\documentclass[11pt,prd,preprint,superscriptaddress]{revtex4}
\usepackage{revsymb}
\usepackage{graphicx}
\usepackage{bm}
\usepackage{amsfonts}
\usepackage{amsmath}
\usepackage{amssymb}
\usepackage{graphicx}

\begin{document}

\title{Power-law solutions and accelerated expansion in scalar-tensor theories}

\author{C.E.M. Batista\footnote{E-mail: cedumagalhaes@hotmail.com}}
\affiliation{Universidade Federal do Esp\'{\i}rito Santo,
Departamento
de F\'{\i}sica\\
Av. Fernando Ferrari, 514, Campus de Goiabeiras, CEP 29075-910,
Vit\'oria, Esp\'{\i}rito Santo, Brazil}

\author{W. Zimdahl\footnote{E-mail: winfried.zimdahl@pq.cnpq.br}}
\affiliation{Universidade Federal do Esp\'{\i}rito Santo,
Departamento
de F\'{\i}sica\\
Av. Fernando Ferrari, 514, Campus de Goiabeiras, CEP 29075-910,
Vit\'oria, Esp\'{\i}rito Santo, Brazil}

\begin{abstract}
We find exact power-law solutions for scalar-tensor theories and clarify the conditions under which they can account for an accelerated expansion of the Universe. These solutions have the property that the signs of both the Hubble rate and the deceleration parameter in the Jordan frame may be different from the signs of their Einstein-frame counterparts.
For special parameter combinations we identify these solutions with asymptotic attractors that have been obtained in the literature through dynamical-system analysis. We establish an effective general-relativistic description for which the geometrical equivalent of dark energy is associated with a time dependent equation of state. The present value of the latter is consistent with the observed cosmological ``constant". We demonstrate that this type of power-law solutions for accelerated expansion cannot be realized in $f(R)$ theories.
\end{abstract}


\maketitle
\date{\today}

\section{Introduction}
\label{Introduction}

The past decade has seen a continuously growing activity, both theoretical and observational, to achieve a consistent picture of the large-scale cosmological dynamics. A huge amount of data has been accumulated which directly or indirectly seem to back up the conclusion, first obtained in \cite{Riess}, that our current Universe entered a phase of accelerated expansion. Direct support is provided by the luminosity-distance data of supernovae of type Ia \cite{SNIa} (but see also \cite{sarkar}), indirect support comes from the anisotropy spectrum of the cosmic microwave background radiation (CMBR) \cite{cmb}, from large-scale-structure data \cite{lss}, from the integrated Sachs--Wolfe effect
\cite{isw}, from baryonic acoustic
oscillations \cite{eisenstein} and from gravitational lensing \cite{weakl}.

A satisfactory explanation of this originally surprising result has not yet been achieved so far.
The component responsible for the accelerated expansion is called dark energy (DE) but its physical nature
remains unclear.
The first guess, an effective cosmological constant, remained the favored option until today and gave rise to the
$\Lambda$CDM model which also plays the role of a standard reference for any research in the field.
According to the interpretation of the data within this model, our Universe is dynamically dominated by a cosmological constant $\Lambda$ which contributes roughly 75\% to the total cosmic energy budget. Roughly 20\% are contributed by cold dark matter (CDM) and only about 5\% are in the form of conventional, baryonic matter. Because of the cosmological constant problem in its different facets, including the coincidence problem (see, e.g., \cite{straumann,pad}), a great deal of work was devoted to alternative approaches in which a similar dynamics as that of the $\Lambda$CDM model is reproduced
with a time varying cosmological term, i.e., the cosmological constant is dynamized.
An important class here are the quintessence models which, in a sense, mimic scalar field inflationary models for the early universe, albeit on a very different energy scale. Also scalar fields with a non-standard kinetic term, called k-essence, have attracted attention.
A further class are unified models of the dark sector, i.e. models, for which a single component plays the role of CDM in the past and the role of a mixture of CDM and DE at the present time.
The generalized Chaplygin gas is the prototype for this kind of models \cite{ugo,berto}. Also viscous fluid models belong to this category \cite{padchi,JCAP}. Common to all these approaches is that Einstein's General Relativity (GR) is assumed to be the valid gravitational theory up to the largest cosmological scales.
But the apparently strange nature of DE and the fact that both dark matter (DM) and DE manifest themselves only gravitationally has also provoked the exploration of alternative theories of gravity.
A major line of investigation in this context are scalar-tensor theories.
Brans-Dicke theory, based on ideas of Mach and Jordan \cite{jordan}, is the prototype of a scalar tensor theory \cite{BD,B}.
According to Mach's principle, the gravitational coupling depends on the mass distribution in the Universe.  Hence the (effective) gravitational ``constant" may be a variable quantity since the mass distribution may vary. Based on this motivation, Brans and Dicke introduced a non-minimally coupled scalar field in order to modify the Newtonian gravitational interaction.
Also more general scalar-tensor theories that have been developed subsequently, are characterized by the fact that the gravitational interaction is mediated both by a metric tensor and a scalar field.
The interest in modified theories of gravity is triggered by the expectation to describe the late-time
accelerated expansion of the Universe without a DE component \cite{carroll,nojiri1,gannouji1}. Instead, it is the geometrical sector which is supposed to provide the desired dynamics which is then gravitationally induced \cite{copeland,torres,lobo,caldkam}.
This class of approaches can be seen as a geometrization of DE.
Fundamental quantum theories involving extra dimension typically seem  to predict 4-dimensional scalar-tensor theories \cite{peri}.
Modified gravitational theories in which the Ricci scalar $R$ in the Einstein-Hilbert action for GR is replaced by a function of $R$ are called $f(R)$ theories. This line of investigation has attracted particular attention as an alternative to GR (\cite{staro,richard,cct,capogrg}).
$f(R)$ theories were shown to be  special cases of scalar-tensor theories \cite{teyssandier,Dwands,chiba}).
With a suitable self-interacting potential and an appropriate coupling to matter fields (chameleon effect) these theories are potential candidates for a geometrical description of DE \cite{chameleon,faulkner,hu,staro2,nojiri} (see also \cite{clifton,narayan}).
Conditions for the cosmological viability of $f(R)$ dark energy models were formulated in \cite{gannouji}.
For recent reviews on the status of these theories see \cite{capofarao,felicetsuji,jainkhoury}.

Scalar-tensor theories are formulated both in the Einstein frame and in the Jordan frame, the two being related by conformal transformations. They are considered to be physically equivalent, although there may occur  differences concerning the General Relativity (GR) limit \cite{kuusk}. On the quantum level the equivalence may be lost. For a discussion of apparent interpretation problems see \cite{catena,farao,CAPOZ}.
Various aspects of scalar-tensor theories in general or subclasses of them have been investigated
\cite{dolgov,chiba,duma,sofarao,soti,brookfield}.
Scalar-tensor theories have also been used in attempts to provide a geometrical explanation of DM \cite{lobo,matos,capo,boehmer,capogrg}.

Scalar-tensor theories are more complex than GR.
Even if the symmetries of the cosmological principle are imposed, scalar-tensor theories do not admit simple solutions that could be compared with, say,  the GR power-law solutions for the cosmic scale factor for fluids with constant equation of state (EoS) parameters.
Investigations in the literature frequently rely on a dynamical system analysis with the aim to find critical points, equivalent to asymptotic power-law solutions \cite{ascale,dunsby,apt,abean,maeda}.
But to the best of our knowledge, exact power-law solution do only exist for special classes of ``curvature quintessence" and without a matter component \cite{capofunaro} but not in the general case.
It is the aim of this paper to partially fill this gap and to derive simple exact solutions which allow for a transparent discussion of at least some of the aspects of these theories.
To this purpose we start looking for solutions with a constant ratio of the energy densities of the matter and the scalar field components in the Einstein frame. The resulting scaling solutions imply a relation between the equation of state (EoS) parameters of the components and the interaction-strength parameter.
For previous studies of scaling solutions of the cosmological dynamics see, e.g.,  \cite{wetterich,uzan,ascale,amendola99,Wetterich,CoLiWands,Ferreira,ZBC}.
The Einstein-frame solutions are then transformed into power-law solutions of the Jordan frame, representing a one-component description of the dynamics that can be confronted with results from dynamical system analysis.
In a next step we interpret the Jordan-frame solutions as solutions of an effective Friedmann equation within GR. Separating a conserved matter part from the total substratum enables us to identify the effective EoS of the (conserved) remaining part that now is regarded as the equivalent of DE. This two-component description is then compared with the $\Lambda$CDM model.

The paper is organized as follows. Section \ref{Basic dynamics} recalls the basic dynamics of scalar-tensor theories both in the Jordan and in the Einstein frames.
In Section \ref{two-component} we obtain scaling solutions in the Einstein frame. The corresponding Jordan-frame solutions are found in Section \ref{Jordan}. Based on these results we discuss the effective EoS of the cosmic medium and point out the relation of our approach to the $\Lambda$CDM model in
Section \ref{cosmed}. In  Section \ref{observations} we test the obtained power-law solution against supernova-type-Ia data. Section \ref{discussion} provides a summary of the paper.



\section{Basic dynamics}
\label{Basic dynamics}

We start by reviewing the basic relations for scalar-tensor theories where we adopt the notation of \cite{abean}.
Scalar-tensor theories are based on the (Jordan-frame) Brans-Dicke type action
\begin{eqnarray}
S &=& \int d^{4}x\sqrt{-g}\left[\frac{1}{2\kappa^{2}}F\left(\Phi\right)R -
\frac{3\left(1 - 4\beta^{2}\right)}{16\kappa^{2}\beta^{2}}\frac{1}{F\left(\Phi\right)}\left(\frac{dF}{d\Phi}\right)^{2}
\left(\nabla \Phi\right)^{2} - V\left(\Phi\right)\right] \nonumber\\ && + \int d^{4}x\sqrt{-g}L_{m}\left(g_{\mu\nu}\right)
 \ ,
 \label{S}
\end{eqnarray}
where $\kappa^{2} = 8\pi G$ and $\beta$ is a coupling constant. The quantities $g$ and $R$ are the determinant and the curvature scalar of the metric tensor $g_{\mu\nu}$, respectively. $L_{m}$ denotes the matter Lagrangian. $F > 0$ is a function of the scalar field $\Phi$ with a potential $V(\Phi)$.
With the help of the transformations
\begin{equation}
g_{\mu\nu} = e^{-2\,\sqrt{2/3}\kappa\beta\phi}\,\tilde{g}_{\mu\nu} \ ,\qquad \
F(\Phi) = e^{2\sqrt{2/3}\kappa\beta\phi}\ ,\qquad \ V(\Phi)= F(\Phi)^{2}\,\tilde{V}(\phi)\ ,
 \ \label{gtil}
\end{equation}
one obtains the Einstein frame action
\begin{equation}
S = \int d^{4}x\sqrt{-\tilde{g}}\left[\frac{1}{2\kappa^{2}}\,\tilde{R} - \frac{1}{2}\left(\tilde{\nabla} \phi\right)^{2} - \tilde{V}\left(\phi\right)
\right] + \int d^{4}x\sqrt{-\tilde{g}}\tilde{L}_{m}\left(e^{-2\sqrt{2/3}\beta\phi}\tilde{g}_{\mu\nu}\right)
 \ .\label{SE}
\end{equation}
An explicit expression for $F(\Phi)$ is required to specify $\Phi$ as a function of $\phi$.
Throughout the paper, quantities with a tilde refer to the Einstein frame, quantities without tilde have their meaning in the Jordan frame.

In the following we restrict our attention to the application of the general theory to homogeneous and isotropic cosmological models with flat spatial sections. Moreover, we assume that the matter part can be modeled by a perfect fluid. Under these conditions the relevant equations are
\begin{equation}
H^{2} = \frac{8\pi G}{3F}\,\left[\rho_{m} +
V(\Phi)\right] - \frac{H}{F}\frac{d F}{dt} +
\frac{1 - 4 \beta^{2}}{16 \beta^{2}}\frac{1}{F^{2}}\left(\frac{d
F}{d t}\right)^{2}
 \ ,
\label{tfried}
\end{equation}
where $H = \frac{1}{a}\frac{da}{dt}$ is the Jordan-frame Hubble rate,
\begin{equation}
\frac{d H}{d t} = - \frac{1}{2F}8 \pi
G\left(\rho_{m} + p_{m}\right) - \frac{1}{2
F}\frac{d^{2}F}{d t^{2}} +
\frac{1}{2}\frac{H}{F}\frac{d F}{dt} - 3 \frac{1 -
4 \beta^{2}}{16 \beta^{2}}\frac{1}{F^{2}}\left(\frac{d F}{d
t}\right)^{2}
 \ ,
\label{dtHtfin}
\end{equation}
\begin{equation}
\frac{d^{2} F}{dt^{2}}  + 3 H\frac{d
F}{dt} = \frac{4}{3} 8 \pi G \beta ^{2}\left[4
V(\Phi) - 2 V_{,\Phi}\,\frac{F}{F_{,\Phi}}
 + \left(\rho_{m} - 3 p_{m}\right)\right] \
\label{ddFfin}
\end{equation}
and
\begin{equation}
\frac{d\rho_{m}}{dt} + 3H
\left(\rho_{m} + p_{m}\right) = 0
 \ .\label{drtm}
\end{equation}
The corresponding Einstein-frame relations are
\begin{equation}
\tilde{H}^{2} = \frac{\kappa^{2}}{3}\left[\tilde{\rho}_{m} + \frac{1}{2}\left(\frac{d\phi}{d\tilde{t}}\right)^{2} + \tilde{V}\right]
 \ ,\label{tH2}
\end{equation}
with the Einstein-frame Hubble rate $\tilde{H} = \frac{1}{\tilde{a}}\frac{d\tilde{a}}{d\tilde{t}}$,
\begin{equation}
\frac{d\tilde{H}}{d\tilde{t}} = - \frac{\kappa^{2}}{2}\left[\tilde{\rho}_{m} + \tilde{p}_{m} + \left(\frac{d\phi}{d\tilde{t}}\right)^{2}
\right]
 \ ,\label{dtH}
\end{equation}
\begin{equation}
 \frac{d^{2}\phi}{d\tilde{t}^{2}} + 3\tilde{H}\frac{d\phi}{d\tilde{t}} + \tilde{V}_{,\phi} = \sqrt{\frac{2}{3}}\kappa\beta\left(\tilde{\rho}_{m} - 3\tilde{p}_{m}\right)
 \ ,\label{ddphi}
\end{equation}
and
\begin{equation}
\frac{d\tilde{\rho}_{m}}{d\tilde{t}} + 3\tilde{H} \left(\tilde{\rho}_{m} + \tilde{p}_{m}\right) =
-\sqrt{\frac{2}{3}}\kappa\beta\frac{d\phi}{d\tilde{t}}\left(\tilde{\rho}_{m} -
3\tilde{p}_{m}\right)
 \ .\label{drm}
\end{equation}
The time coordinates and the scale factors of the Robertson-Walker metrics of both frames are related by
\begin{equation}
dt =
e^{-\,\sqrt{2/3}\kappa\beta\phi} d \tilde{t}\ \qquad \mathrm{and} \qquad a =
e^{-\,\sqrt{2/3}\kappa\beta\phi}\tilde{a} \ ,\label{ttil}
\end{equation}
respectively.
The matter quantities transform into each other via
\begin{equation}
p_{m} = e^{4\,\sqrt{2/3}\kappa\beta\phi}\,\tilde{p}_{m} \
,\qquad \rho_{m} = e^{4\,\sqrt{2/3}\kappa\beta\phi}\,\tilde{\rho}_{m}\ .\label{mtil}
\end{equation}
This means $\frac{p_{m}}{\rho_{m}} = \frac{\tilde{p}_{m} }{\tilde{\rho}_{m}}$, i.e., the
EoS parameter remains invariant.

\section{A two-component description}
\label{two-component}

\subsection{General relations}

The set of Einstein-frame equations (\ref{tH2}) -- (\ref{drm}) suggests an effective two-component structure in which matter interacts with a scalar field.
One may attribute an effective energy density $\tilde{\rho}_{\phi}$ and an effective pressure $\tilde{p}_{\phi}$ to the scalar field by
\begin{equation}
\tilde{\rho}_{\phi} = \frac{1}{2}\left(\frac{d\phi}{d\tilde{t}}\right)^{2}+ \tilde{V}\qquad \mathrm{and} \qquad \tilde{p}_{\phi} =
\frac{1}{2}\left(\frac{d\phi}{d\tilde{t}}\right)^{2} - \tilde{V}
 \ ,\label{rhophi}
\end{equation}
respectively.
Equations (\ref{drm}) and (\ref{ddphi}) can be written as
\begin{equation}
\frac{1}{\tilde{\rho}_{m}}\frac{d\tilde{\rho}_{m}}{d\tilde{t}}  + 3\tilde{H} \left(1 + \tilde{w}_{m}\right) =
-\sqrt{\frac{2}{3}}\kappa\beta\frac{d\phi}{d\tilde{t}}\left(1- 3\tilde{w}_{m}\right)
 \ \label{drm+}
\end{equation}
and
\begin{equation}
\frac{1}{\tilde{\rho}_{\phi}}\frac{d\tilde{\rho}_{\phi}}{d\tilde{t}}  + 3\tilde{H} \left(1 +
\tilde{w}_{\phi}\right) = \sqrt{\frac{2}{3}}\kappa\beta\frac{d\phi}{d\tilde{t}}\left(1 -
3\tilde{w}_{m}\right)r
 \ ,\label{drp+}
\end{equation}
respectively, where we have introduced the matter EoS parameter $\tilde{w}_{m} =
\frac{\tilde{p}_{m}}{\tilde{\rho}_{m}}$ and the Einstein-frame EoS parameter for the scalar field  $\tilde{w}_{\phi} = \frac{\tilde{p}_{\phi}}{\tilde{\rho}_{\phi}}$.
Furthermore, we
have defined the ratio of the energy densities
\begin{equation}
r \equiv \frac{\tilde{\rho}_{m}}{\tilde{\rho}_{\phi}}
 \ .\label{r}
\end{equation}
The total energy density and the total pressure are
\begin{equation}
\tilde{\rho} = \tilde{\rho}_{m} + \tilde{\rho}_{\phi}\ ,\qquad  \tilde{p} = \tilde{p}_{m} + \tilde{p}_{\phi} \quad
\Rightarrow\quad \frac{d\tilde{\rho}}{d\tilde{t}} + 3\tilde{H} \left(\tilde{\rho} + \tilde{p}\right) = 0
 \ .\label{rho}
\end{equation}
The set of equations (\ref{drm+}) and (\ref{drp+}) is reminiscent of interacting quintessence models (see, e.g.,~\cite{ZBC}), in which the scalar field is part of the energy-momentum tensor within standard GR and interacts with the matter component in a specific way.
From this point of view (\ref{drm+}) and (\ref{drp+}) can be regarded as a pure GR-based model. Then the parameter $r$ describes the ratio of dark matter to quintessential dark energy. However, such type of interpretation masks the circumstance that in the present context the scalar field is a gravitational degree of freedom.
But the formal equivalence between (\ref{drm+}) and (\ref{drp+}) and corresponding equations in GR may be used to apply solution techniques of the latter to the former situation.
This will exactly be our strategy in the following subsection. A physical interpretation will be given subsequently within the Jordan frame.

We mention that the coupling parameter $\beta$ is related to the Brans-Dicke parameter $\omega_{BD}$ by
\begin{equation}
\omega_{BD} = \frac{3}{8}\,\frac{1 - 4 \beta^{2}}{\beta^{2}} \quad \Leftrightarrow\quad
\beta^{2} = \frac{3}{4}\,\frac{1}{2\omega_{BD} + 3}
 \ .\label{obd}
\end{equation}
It is known that $f(R)$ theories can be regarded as a subclass of scalar-tensor theories, corresponds to the special case $\beta^{2} = \frac{1}{4}$ in Eq.~(\ref{S}) (cf. \cite{teyssandier,Dwands,chiba}), equivalent to $\omega_{BD} = 0$. These theories seemed to be ruled out observationally \cite{dolgov,chiba}, but according to \cite{chameleon,faulkner,hu,staro2,nojiri,felicetsuji,jainkhoury}, a suitable effective potential may result in a sufficiently heavy  mass of the scalar field in regions of high matter density (chameleon mechanism), so that conflicts with solar system constraints can be avoided. The circumstance that scalar-tensor theories are related to a non-linear Lagrangian by a conformal transformation is also known as Bicknell-theorem \cite{bicknell,hjschmi}.

\subsection{Scaling solutions}

From now on we shall focus on the subclass of solutions for the system (\ref{drm+}) and (\ref{drp+}) that admit a constant ratio of the energy densities of both components. Corresponding solutions in GR are of interest in connection with the coincidence problem, i.e. the question, why the densities of DM and DE are of the same order just at the present time. In the present context such connection is less obvious since for an adequate interpretation one has also to consider the Jordan-frame solutions which will be found below in section \ref{Jordan}. The general dynamics of the energy density ratio $r$ is obtained by
differentiating (\ref{r}) and using the balances (\ref{drm+}) and (\ref{drp+}):
\begin{equation}
\frac{dr}{d\tilde{t}} = 3\tilde{H} r\left[\tilde{w}_{\phi} - \tilde{w}_{m} - \sqrt{\frac{2}{3}}\kappa\beta \frac{1}{3 \tilde{H}}\frac{d\phi}{d\tilde{t}}\left(1 - 3\tilde{w}_{m}\right)\left(1 +
r\right)\right]
 \ .\label{dr+}
\end{equation}
Scaling solutions, i.e. solutions with a constant value of the Einstein-frame energy density ratio $r$,
are then characterized by
\begin{equation}
\frac{dr}{d\tilde{t}} = 0 \quad \Leftrightarrow\quad
\sqrt{\frac{2}{3}}\kappa\beta\frac{d\phi}{d\tilde{t}}\left(1 - 3\tilde{w}_{m}\right) = 3\tilde{H}
\,\frac{\tilde{w}_{\phi} -
\tilde{w}_{m}}{1 + r}
 \ .\label{dr0b}
\end{equation}
Notice that $\tilde{w}_{m} = \frac{1}{3}$ corresponds to the interaction-free case in (\ref{drm+}) and (\ref{drp+}). Under this condition a scaling solution can only exist if $\tilde{w}_{\phi} = \frac{1}{3}$ as well.
The left-hand side of the second equation (\ref{dr0b}) determines the source (loss) terms on the right-hand sides of the balance equations (\ref{drm+}) and (\ref{drp+}).
Consequently, the latter equations become
\begin{equation}
\frac{d\tilde{\rho}_{\phi}}{d\tilde{t}}  + 3\tilde{H} \left(\tilde{\rho}_{\phi} + \tilde{p}_{\phi}\right) =
3\tilde{H}
\,\frac{\tilde{w}_{\phi} -
\tilde{w}_{m}}{1 + r}\,r\tilde{\rho}_{\phi}
 \ .\label{drpfin}
\end{equation}
and
\begin{equation}
\frac{d\tilde{\rho}_{m}}{d\tilde{t}}  + 3\tilde{H} \left(\tilde{\rho}_{m} + \tilde{p}_{m}\right) = -
3\tilde{H}
\,\frac{\tilde{w}_{\phi} -
\tilde{w}_{m}}{1 + r}\,\tilde{\rho}_{m}
 \ ,\label{drmfin}
\end{equation}
respectively.
Hence, for scaling solutions the coupling is completely specified. No free parameter occurs on the right-hand sides of Eqs.~(\ref{drpfin}) and~(\ref{drmfin}).  The direction of the energy flow depends on the sign of
$\tilde{w}_{\phi} -\tilde{w}_{m}$. For $\tilde{w}_{\phi} -\tilde{w}_{m} > 0$ we have a flux from the $m$ component to the $\phi$
component. For $\tilde{w}_{\phi} -\tilde{w}_{m} < 0$ it is the opposite (always assuming $\tilde{H} > 0$).

While the set of equations (\ref{drm+}) and (\ref{drp+}) is well known  in the literature, the configuration (\ref{drpfin}) and~(\ref{drmfin}) does not seem to have attracted attention so far. Studying the system (\ref{drpfin}) and~(\ref{drmfin}), exploring its consequences for the Jordan frame
and discussing implications for the corresponding cosmological dynamics are the main aims of this paper.

For constant values of  $\tilde{w}_{m}$ and $\tilde{w}_{\phi}$ equations (\ref{drpfin}) and~(\ref{drmfin}) have the solutions
\begin{equation}
\tilde{\rho}_{\phi} \propto \tilde{\rho}_{m} \propto  \tilde{a}^{-3\,\frac{1 + \tilde{w}_{\phi} + r\left(1 +
\tilde{w}_{m}\right)}{1 + r}} \ .\label{rp+}
\end{equation}
Here we have used that integrating Eq.~(\ref{dr0b}) yields
$
e ^{\sqrt{2/3}\kappa\beta\,\phi} \propto
\tilde{a}^{-3\,\frac{\tilde{w}_{m} - \tilde{w}_{\phi}}{\left(1 -
3\tilde{w}_{m}\right)\left(1 + r\right)}}$.
The Friedmann equation (\ref{tH2}) takes the form
$
3\tilde{H}^{2} = 8 \pi G \tilde{\rho} = 8 \pi G \left(1 + r\right)\tilde{\rho}_{\phi}
$
which results in
\begin{equation}
\tilde{a} \propto \tilde{t}^{\frac{2}{3}\frac{1 + r}{1 + \tilde{w}_{\phi} +
r\left(1 + \tilde{w}_{m}\right)}} \
 \ \label{a/ai2}
\end{equation}
for the scale factor.
Obviously, the correct limits for $r
\rightarrow 0$ and $r \rightarrow \infty$ are consistently
recovered.
The dynamics is that of a substratum with an effective EoS
\begin{equation}
\tilde{w}_{\mathrm{eff}} = \frac{ \tilde{w}_{\phi} + r \tilde{w}_{m}}{1 + r} \quad \Rightarrow\quad
\tilde{a} \propto \tilde{t}^{\frac{2}{3 \left(1 + \tilde{w}_{\mathrm{eff}} \right)}} \ ,\qquad
\tilde{H} = \frac{2}{3 \left(1 + \tilde{w}_{\mathrm{eff}} \right)}\,\frac{1}{\tilde{t}}
\ .
 \ \label{aHeff}
\end{equation}
There exists also a contracting solution
\begin{equation}
\tilde{a} = \tilde{a}_{i}\left(\frac{\tilde{t}_{f} - \tilde{t}}{\tilde{t}_{f} - \tilde{t}_{i}}\right)^{\frac{2}{3 \left(1 + \tilde{w}_{\mathrm{eff}} \right)}} \ ,\qquad
\tilde{H} = -\frac{2}{3 \left(1 + \tilde{w}_{\mathrm{eff}} \right)}\,\frac{1}{\tilde{t}_{f} -\tilde{t}}
\ ,
\label{aHeffc}
\end{equation}
where $\tilde{t}_{i}, \tilde{t} < \tilde{t}_{f}$.
With (\ref{aHeff}) (or (\ref{aHeffc})) the Einstein-frame dynamics is completely solved for the given configuration. The similarity to GR power-law solutions is obvious. But as already mentioned, we consider the solutions (\ref{aHeff}) here as an intermediate result and postpone a physical discussion to the next sections.

For the deceleration parameter $\tilde{q} \equiv - \frac{1}{\tilde{a}\tilde{H}^{2}}
\frac{d^{2}\tilde{a}}{d\tilde{t}^{2}} = - 1 - \frac{1}{\tilde{H}^{2}}\frac{d\tilde{H}}{d\tilde{t}}$
we obtain
$
\tilde{q} = \frac{1}{2}\,\left(1 + 3\tilde{w}_{\mathrm{eff}}\right)
$.
The condition for accelerated expansion in the Einstein frame is
\begin{equation}
\tilde{w}_{\mathrm{eff}} < -\frac{1}{3} \quad \Leftrightarrow\quad \tilde{w}_{\phi} + r \tilde{w}_{m} <
-\frac{1}{3}\left(1 + r\right)\ .\label{dda}
\end{equation}
An accelerated expansion can be obtained for $\tilde{w}_m = 0$ and $\tilde{w}_{\phi} < -\frac{1}{3}\left(1 + r\right)$.
The special case of exponential expansion is characterized by
\begin{equation}
\tilde{w}_{\mathrm{eff}} = - 1 \quad
\Leftrightarrow\quad 1 + \tilde{w}_{\phi} + r\left(1 + \tilde{w}_{m}\right) = 0 \quad
\Leftrightarrow\quad \tilde{w}_{\phi} + r \tilde{w}_{m} = -1 - r
 \ .\label{expex}
\end{equation}
Notice that $\tilde{w}_{\phi} \geq -1$. Consequently,
\begin{equation}
 \tilde{w}_{m} = -1 - \frac{1 + \tilde{w}_{\phi}}{r}
 \ ,\label{expexw}
\end{equation}
i.e., $ \tilde{w}_{m} \leq -1$, the component $m$ has to be of the phantom type.
A standard matter type solution $\tilde{\rho} \propto \tilde{a}^{-3}$ is obtained for $\tilde{w}_{\mathrm{eff}} = 0$.
According to (\ref{aHeff}), this is realized for any combination  $\tilde{w}_{\phi} + r \tilde{w}_{m} = 0$.

From Eqs.~(\ref{dr0b}) and Friedmann's equation it follows that
\begin{equation}
\frac{1}{2}\left(\frac{d\phi}{d\tilde{t}}\right)^{2} = \frac{9}{4}\frac{\left(\tilde{w}_{m} -
\tilde{w}_{\phi}\right)^{2}}{\beta^{2}\left(1 -
3\tilde{w}_{m}\right)^{2}\left(1+r\right)}\,\tilde{\rho}_{\phi} \ .\label{dp2}
\end{equation}
The potential is given by
\begin{equation}
\tilde{V}\left(\phi\right) = \tilde{\rho}_{\phi} - \frac{1}{2}\left(\frac{d\phi}{d\tilde{t}}\right)^{2} =
\left[1 - \frac{9}{4}\frac{\left(\tilde{w}_{m} -
\tilde{w}_{\phi}\right)^{2}}{\beta^{2}\left(1 -
3\tilde{w}_{m}\right)^{2}\left(1+r\right)}\right]\,\tilde{\rho}_{\phi}
 \ .\label{V}
\end{equation}
Consistency between (\ref{V}) and $\tilde{V} = \frac{1}{2}\left(1-\tilde{w}_{\phi}\right)\tilde{\rho}_{\phi}$ implies the following expression for the interaction parameter,
\begin{equation}
\beta^{2} = \frac{9}{2}\frac{\left(\tilde{w}_{m}-\tilde{w}_{\phi}\right)^{2}}{\left(1-3\tilde{w}_{m}\right)^{2}}
\frac{1}{1+r}\cdot\frac{1}{1+\tilde{w}_{\phi}}
 \ .\label{beta=}
\end{equation}
As already mentioned, the interaction constant in the balances (\ref{drm+}) and (\ref{drp+}) is not
a free parameter for the given configuration, but determined by the equation of state parameters
and the ratio $r$.
Eliminating $\beta$ via (\ref{beta=}), the potential may be written as
\begin{equation}
\tilde{V}\left(\phi\right) = \tilde{V}_{i}\, \exp\{-\sqrt{24\pi G}\sqrt{\frac{1 + r}{1 + \tilde{w}_{\phi}}}\,\left(1 + \tilde{w}_{\mathrm{eff}}\right)\, \left(\phi
-\phi_{i}\right)\}
 \ .\label{Vp3}
\end{equation}
With  (\ref{Vp3}) and (\ref{dr0b}) the scalar-field equation (\ref{ddphi})
is identically satisfied.
Finally we notice that
combination of (\ref{obd}) and (\ref{beta=}) results in
\begin{equation}
\omega_{BD} = - \frac{3}{2} + \frac{1}{12}\,\frac{\left(1 - 3\tilde{w}_{m}\right)^{2}}{\left(\tilde{w}_{m} - \tilde{w}_{\phi}\right)^{2}}\left(1 + r\right)\left(1 + \tilde{w}_{\phi}\right)
 \ \label{obd=}
\end{equation}
for the Brans-Dicke parameter. Accelerating solutions for dust with $\omega_{BD} < 0$ have been studied in \cite{julio,diego2}.

This concludes our consideration of scaling solutions in the Einstein frame. In the following subsection we shall use the transformations (\ref{ttil}) to obtain the corresponding Jordan-frame dynamics.  While this dynamics will also be characterized by power-law solutions, it is not associated with a constant ratio of the energy densities of the dynamically relevant components.

\section{Jordan frame solutions}
\label{Jordan}
\subsection{General power-law structure}
The transformations from Einstein's to Jordan's frame are mediated by exponentials of $\phi$.
Via relations (\ref{ttil}) one finds
\begin{equation}
\frac{d a}{dt} = \frac{d \tilde{a}}{d\tilde{t}} - \sqrt{2/3}\kappa\beta \frac{d \phi}{d\tilde{t}} \tilde{a} \ .\label{dtila}
\end{equation}
The Hubble rates are related by
\begin{equation}
H \propto  \frac{1 -
3\tilde{w}_{\mathrm{eff}}}{1 -
3\tilde{w}_{m}}\,\tilde{a}^{-3\,\frac{\tilde{w}_{m} - \tilde{w}_{\mathrm{eff}}}{\left(1 -
3\tilde{w}_{m}\right)}}\,
 \tilde{H}
 \ .\label{tilHH}
\end{equation}
This means, the Hubble rates of both frames do not necessarily have the same sign.
For $\tilde{w}_{m} = 0$ and $\tilde{w}_{\mathrm{eff}} > \frac{1}{3}$, e.g., an expanding solution in the Jordan frame corresponds to a contracting solution in the Einstein frame. 
Similar properties of conformally related frames have been discussed in string-theory based pre-big bang scenarios \cite{gasperini}. 
The explicit relations between the scale factors $a$ and $\tilde{a}$ and the time coordinates $t$ and
$\tilde{t}$  are
\begin{equation}
a \propto
\tilde{a}^{\frac{1 - 3\tilde{w}_{\mathrm{eff}}}{1 - 3\tilde{w}_{m}}}\ \quad\ \mathrm{and} \quad \
t\ \propto \
\tilde{t}^{\ \frac{1 - \tilde{w}_{\mathrm{eff}} -\tilde{w}_{m}\left(1 + 3\tilde{w}_{\mathrm{eff}}\right)}{\left(1 -
3\tilde{w}_{m}\right)\left(1 + \tilde{w}_{\mathrm{eff}}\right)}}
,\label{tilaa}
\end{equation}
respectively. These relations encode the correspondence between Einstein frame dynamics and Jordan frame dynamics for our power-law solutions.
Combination with  (\ref{aHeff}) provides us with the result
\begin{equation}
a\ \propto \  t^{\ \frac{2}{3\left(1 + w_{\mathrm{eff}}\right)}}\ \quad\ \mathrm{with} \quad \
w_{\mathrm{eff}} = \frac{2\tilde{w}_{\mathrm{eff}} - \tilde{w}_{m}\left(1 + 3 \tilde{w}_{\mathrm{eff}}\right)}
{1 - 3 \tilde{w}_{\mathrm{eff}}}
\ .\label{tatt}
\end{equation}
With (\ref{tatt}) the Jordan-frame dynamics is solved as well.
By direct calculation one checks that with the Jordan-frame solution (\ref{tatt}) the set of equations
(\ref{tfried}) - (\ref{drtm}) is satisfied.
The solution is of the power-law type, corresponding to a constant effective EoS parameter. It is $w_{\mathrm{eff}}$ which physically characterizes the EoS of the cosmic substratum. Obviously, the parameter $w_{\mathrm{eff}}$ differs from the Einstein-frame parameter $\tilde{w}_{\mathrm{eff}}$, the letter being an auxiliary quantity.
While constant EoS parameters certainly cannot account for a continuous transition between different epochs of the cosmological evolution, such as the transition from  matter to  dark-energy type dominance, they are useful as simple exact solutions which are valid piecewise.
One may associate a conserved total effective energy density
\begin{equation}
\rho_{\mathrm{eff}} \propto a^{-3\left(1 + w_{\mathrm{eff}}\right)}
\ \label{reff}
\end{equation}
to the solution (\ref{tatt}).
The introduction of this quantity allows us to understand our solution alternatively as a GR solution of an effective Friedmann equation
$3H^{2} = 8 \pi G \rho_{\mathrm{eff}}$
for a medium with EoS parameter $w_{\mathrm{eff}}$.
We shall come back this point in section \ref{cosmed}.

The Jordan-frame deceleration parameter can be written as
\begin{equation}
q \equiv -
\frac{a\frac{d^{2}a}{dt^{2}}}
{\left(\frac{da}{dt}\right)^{2}} \qquad \Rightarrow\qquad
q = \frac{1}{2}\left(1 + 3 w_{\mathrm{eff}}\right)
\ ,\label{tilqeff}
\end{equation}
The relation to the Einstein-frame deceleration parameter is
\begin{equation}
q = \tilde{q} \frac{\left(1 - 3\tilde{w}_{m}\right)\left(1+r\right)}{1 - 3
\tilde{w}_{\phi} + r\left(1 - 3\tilde{w}_{m}\right)} = \tilde{q} \frac{1 - 3\tilde{w}_{m}}{1 - 3
\tilde{w}_{\mathrm{eff}}}
 \  .\label{tilqq}
\end{equation}
Notice that also $q$ and $\tilde{q}$ do not necessarily have the same sign.
For the already mentioned case $\tilde{w}_{m} = 0$ and $\tilde{w}_{\mathrm{eff}} > \frac{1}{3}$ deceleration in the Einstein frame corresponds to acceleration in the Jordan frame.
For Einstein-frame exponential expansion (\ref{expex}) it follows that
\begin{equation}
q = - \frac{1}{4}\left(1 - 3\tilde{w}_{m}\right)\ , \qquad\qquad
(\tilde{w}_{\phi} + r \tilde{w}_{m} = -1 - r)\ .\label{tilqexp}
\end{equation}
For the degenerate case $\tilde{w}_{m} = -1$ we obtain $q = -1$ as expected.

\subsection{Solutions for non-relativistic matter $w_{m} = 0$}
 For the particularly interesting case  $w_{m} = 0$ (recall that $\tilde{w}_{m} = w_{m}$), the second relation (\ref{tatt}) reduces to
\begin{equation}
w_{\mathrm{eff}} = \frac{2\tilde{w}_{\mathrm{eff}}}
{1 - 3 \tilde{w}_{\mathrm{eff}}}
\ .\label{tilw0}
\end{equation}
Given the EoS of pressureless matter for the material content of the Universe, Eq.~(\ref{tilw0}) defines the potentially possible physical EoS parameters for the total cosmological dynamics in terms of the Einstein-frame parameters.
The relation between different intervals of $\tilde{w}_{\mathrm{eff}}$ and $w_{\mathrm{eff}}$ is shown in Tab.~\ref{range}.
Interesting special cases in the Jordan frame are:

\noindent
(i) \textit{a matter dominated phase} $w_{\mathrm{eff}} = 0$, realized for $\tilde{w}_{\mathrm{eff}}=0$.
Obviously, this solution requires either  $\tilde{w}_{\phi} = 0$ as well or $r \rightarrow \infty$ (cf. relation (\ref{aHeff})). This phase corresponds to the limit between the ranges
III and IV in Tab.~\ref{range}.
According to
(\ref{beta=}), this matter dominated solution has $\beta^{2}=0$.

\noindent
(ii) \textit{a phase of exponential expansion}, realized for $w_{\mathrm{eff}} = - 1$  in (\ref{tilw0}).
This requires
\begin{equation}
\tilde{w}_{\mathrm{eff}} = 1 \quad \Rightarrow\quad  \tilde{w}_{\phi} = 1 + r
\ \label{w+}
\end{equation}
and corresponds to the limit between I and II in Tab.~\ref{range}.
We emphasize that $w_{\mathrm{eff}} = - 1$ implies $\tilde{w}_{\mathrm{eff}} = + 1$.
While the expansion in the Jordan frame is exponentially accelerated, we have a decelerated contraction in the Einstein frame. Recall (cf. Eq.~(\ref{tilHH})) that for $\tilde{w}_{\mathrm{eff}} > \frac{1}{3}$ the Hubble rates in both frames have different signs. We notice also, that $\tilde{w}_{\phi} > 1$ implies a negative potential $\tilde{V}(\phi)$.

\noindent
(iii) \textit{the onset of accelerated expansion} $w_{\mathrm{eff}} = - \frac{1}{3}$. For this case at the limit between the intervals IV and V in Tab.~\ref{range}, we have $\tilde{w}_{\mathrm{eff}} = - \frac{1}{3}$ as well, with $\tilde{w}_{\phi} = - \frac{1}{3}\left(1+r\right)$.

\noindent
(iv) \textit{an intermediate case}, belonging to range I in Tab.~\ref{range}, is $w_{\mathrm{eff}} = - \frac{3}{4}$, corresponding to $\tilde{w}_{\mathrm{eff}} = 3$ with $\tilde{w}_{\phi} = 3\left(1+r\right)$.

\noindent
(v) \textit{a phantom EoS} $w_{\mathrm{eff}} = - \frac{3}{2}$, part of range II in Tab.~\ref{range}. This requires $\tilde{w}_{\mathrm{eff}} = \frac{3}{5}$.

The existence of power-law solutions with $w_{\mathrm{eff}} < - \frac{1}{3}$ in a universe filled with pressureless matter is possibly the simplest demonstration for the capability of scalar-tensor theories to account for an accelerated expansion without a dark-energy component. To the best of our knowledge a power-law behavior is known so far only asymptotically but not as an exact solution of the full theory.

In a next step we investigate, which EoS parameters $w_{\mathrm{eff}}$ in (\ref{tilw0}) are admitted for
$\beta^{2} = \frac{1}{4}$, i.e., for $f(R)$-type theories. With $\tilde{w}_{m}=0$ we find from (\ref{beta=}) and (\ref{aHeff}) that $\beta^{2} = \frac{1}{4}$ is realized for
\begin{equation}
w_{eff} = \frac{2}{3}\,\frac{1 \pm \sqrt{1 + \frac{72}{1+r}}}{11 \mp \sqrt{1 + \frac{72}{1+r}}}
\ .\label{b=1/2}
\end{equation}
\noindent
(vi) $\beta^{2} = \frac{1}{4}$ and $r=0$. The solutions are $w_{eff} \approx 2.5$ and $w_{eff} \approx - 0.26$. Accelerated expansion is impossible under these conditions.

\noindent
(vii) $\beta^{2} = \frac{1}{4}$ and $r \rightarrow\infty$. For very large values of $r$ ($r\gtrsim 10^{3}$) the solutions of (\ref{b=1/2}) are $w_{eff} \approx \frac{2}{15}$ and $w_{eff} \approx 0$. The latter reproduces a matter era with $a \propto t^{2/3}$.

\noindent
(viii) $\beta^{2} = \frac{1}{4}$ and $r = 8$. The upper sign of (\ref{b=1/2}) yields $w_{eff} = \frac{1}{3}$, corresponding to $a \propto t^{1/2}$. This solution has played a role in the discussion about the cosmic viability of $f(R)$ theories \cite{apt}. For the lower sign it follows that  $w_{eff} = - \frac{2}{21}$.
Consequently, there do not exist scaling solutions of the type discussed here in $f(R)$ theories that can describe an accelerated expansion of the Universe.

\subsection{Solutions for $w_{m} \neq 0$}
So far we have considered special cases with $w_{m} = 0$.
But it is also possible to realize  equations of state of interest under the conditions $w_{m} \neq 0$ and   $\tilde{w}_{\phi} = 0$. For $\tilde{w}_{\phi} = 0$ the general EoS parameter in (\ref{tatt}) specifies to
\begin{equation}
w_{\mathrm{eff}} = w_{m}\frac{r\left(1 - 3w_{m}\right) -1}{r\left(1 - 3w_{m}\right) +1}
\ .\label{tilwm}
\end{equation}
This leads to the following quadratic equation for $w_{m}$,
\begin{equation}
w_{m}^{2} + \frac{1}{3}\frac{1 - r\left(1+ 3w_{eff}\right)}{r} w_{m} + \frac{1+r}{3r}w_{eff} = 0
\ .\label{wm2}
\end{equation}
We have

\noindent
(ix)\textit{ a matter-dominated universe $w_{\mathrm{eff}}=0$}. There are two solution, $w_{m}=0$ and $w_{m}=\frac{1}{3}\frac{r-1}{r}$. For $r \gg 1$ the latter approaches $w_{m}=\frac{1}{3}$, i.e., the EoS for radiation.

\noindent
(x) \textit{exponentially accelerated expansion $w_{\mathrm{eff}}=-1$}. In this case the EoS parameter of the matter component is given by
\begin{equation}
w_{m} = - \frac{1}{6}\frac{1+2r}{r}\left[1 \pm \sqrt{1 + 12 r\frac{1+r}{\left(1+2r\right)^{2}}}\right]
\ .\label{wm-1}
\end{equation}
For $r \gg 1$ the solutions are $w_{m}=-1$ and $w_{m}=\frac{1}{3}$. For $r=1$ one has $w_{m}=-1.46$ and $w_{m}=0.45$.
For the solutions with $w_{m} < 0$ it can be argued that now the $\phi$ component describes the (non-relativistic) matter and the $m$ component plays the role of dark energy. Different to the cases in which the dark energy is geometrized, now it is the dark matter.
But again, there exist solutions for an accelerated expansion which do not require negative EoS parameters of the matter component, i.e., solutions with $w_{\mathrm{eff}}=-1$ that have $\tilde{w}_{\phi} = 0$ and $w_{m}>0$.

\subsection{Relation to other work}
In a next step we clarify the relation of our solution (\ref{tatt}) with results from the analysis of dynamical systems. The latter typically determines critical points of the cosmic evolution, which (in some studies) amounts to finding asymptotic power law solutions (see \cite{ascale,dunsby,apt,abean,maeda}). It is therefore interesting to compare the asymptotic power-law behavior obtained within the context of dynamical systems with our exact power-law solutions.
As an example we show the consistency of our solution with several of the critical points found in \cite{abean}.

\noindent
Point P1 in \cite{abean} has $w_{\mathrm{eff}} = -1$ and a vanishing fractional matter contribution,  $\frac{\rho_{m}}{\rho_{\mathrm{eff}}} = 0$. It corresponds to our case (ii) above with $r=0$.

\noindent
Point P2 in \cite{abean} has $w_{\mathrm{eff}} = \frac{8 \beta^{2}}{9 - 12 \beta^{2}}$ and
$\frac{\rho_{m}}{\rho_{\mathrm{eff}}} = \frac{9 - 4\beta^{2}}{\left(3 - 4\beta^{2}\right)^{2}}$
. For $\beta^{2} = 0$ it describes the matter dominated phase of case (i) above. But for $\beta^{2} = \frac{9}{4}$
we can also reproduce $w_{\mathrm{eff}} = -1$, realized for $w_{m}= r= 0$ and $\tilde{w}_{\mathrm{eff}} =\tilde{w}_{\phi} = 1$, corresponding to case (ii) above.

\noindent
Point P3 in \cite{abean} has $w_{\mathrm{eff}} = \frac{3 -2\beta}{3 + 6\beta}$ and $\frac{\rho_{m}}{\rho_{\mathrm{eff}}} = 0$. For our solution (\ref{beta=}) for $\beta$
with $w_{m}= r= 0$ and $\tilde{w}_{\mathrm{eff}} =\tilde{w}_{\phi}$ this can be written
\begin{equation}
w_{\mathrm{eff}} = \frac{3 \mp 3\sqrt{2}
\frac{\tilde{w}_{\mathrm{eff}}}{\sqrt{1 + \tilde{w}_{\mathrm{eff}}}}}{3 \pm 9\sqrt{2}\frac{\tilde{w}_{\mathrm{eff}}}{\sqrt{1 + \tilde{w}_{\mathrm{eff}}}}}
\ .\label{wp3}
\end{equation}
For $\tilde{w}_{\mathrm{eff}}=1$ there is a solution $w_{\mathrm{eff}}=0$ (upper signs) and another one  (lower signs), $w_{\mathrm{eff}}=-1$, which again reduces to our previous case (ii).

\noindent
Finally, Point P7 in \cite{abean} has $w_{\mathrm{eff}} = \frac{3 + 2\beta}{3 - 6\beta}$ and $\frac{\rho_{m}}{\rho_{\mathrm{eff}}} = 0$. Here, $w_{\mathrm{eff}} = -1$ can be obtained for $\beta=\frac{3}{2}$. Again, this requires $\tilde{w}_{\phi} = 1$, i.e., a vanishing potential term.
We conclude that our solution (\ref{tatt}), in particular the special case (\ref{tilw0}), is consistent with results from dynamical system analysis.

\begin{table}[!t]
\begin{center}
\begin{tabular}{|c|c|c|}
  \hline
  $$  & EoS Einstein frame & EoS Jordan frame \\
  \hline
  I & $\infty > \tilde{w}_{\mathrm{eff}} \geq 1 $& $- \frac{2}{3} > w_{\mathrm{eff}} \geq - 1 $\\
  \hline
  II & $1 \geq \tilde{w}_{\mathrm{eff}} > \frac{1}{3}$&$ - 1 \geq w_{\mathrm{eff}} > - \infty $\\
  \hline
  III &$ \frac{1}{3} > \tilde{w}_{\mathrm{eff}} \geq 0$&$ \infty > w_{\mathrm{eff}} \geq 0 $\\
  \hline
  IV & $0 \geq \tilde{w}_{\mathrm{eff}} \geq - \frac{1}{3}$&$ 0 \geq w_{\mathrm{eff}} \geq - \frac{1}{3} $\\
  \hline
  V &$ - \frac{1}{3} \geq \tilde{w}_{\mathrm{eff}} \geq - 1$&$ - \frac{1}{3} \geq w_{\mathrm{eff}} \geq - \frac{1}{2} $\\
  \hline
\end{tabular}
\end{center}
\caption{Correspondence between different ranges for the Einstein-frame EoS parameter $\tilde{w}_{\mathrm{eff}}$ and the corresponding EoS parameter $w_{\mathrm{eff}}$ in the Jordan frame for $w_{m} = 0$.}  \label{range}
\end{table}

\section{The cosmic medium}
\label{cosmed}

Now we come back to the effective  Friedmann equation with an energy density (\ref{reff}). Obviously, (\ref{tatt}) is a solution of Friedmann's equation $3H^{2} = 8 \pi G \rho_{\mathrm{eff}}$. The energy density (\ref{reff}), on the other hand, can be seen as the solution of a conservation equation
\begin{equation}
\frac{d\rho_{\mathrm{eff}}}{dt} + 3 H\left(\rho_{\mathrm{eff}} + p_{\mathrm{eff}}\right) = 0
 \ \label{conseff}
\end{equation}
with an effective pressure
\begin{equation}
p_{\mathrm{eff}} = w_{\mathrm{eff}}\rho_{\mathrm{eff}}
 \ .\label{peff}
\end{equation}
This set of equations represents a GR equivalent for the power-law solution of the scalar-tensor theory.
The cosmic medium as a whole is described by an EoS parameter, given by the second relation in  (\ref{tatt}).
Moreover, we know that a separately conserved matter component with an EoS parameter $w_{m}$ is part of the cosmic medium.
By direct calculation one confirms that upon using (\ref{mtil})
together with the solution (\ref{rp+}) the
conservation relation (\ref{drtm}) is consistently satisfied, corresponding to
\begin{equation}
\rho_{m} \propto a^{-3\left(1 + w_{m}\right)}
 \ .\label{drtm2}
\end{equation}
Let us define an energy density $\rho_{x}$ as the difference
\begin{equation}
\rho_{x} \equiv \rho_{\mathrm{eff}}  - \rho_{m}
 \ .\label{rdiff}
\end{equation}
This means, we assume the total effective energy density $\rho_{\mathrm{eff}}$ to be the sum of a matter contribution and this newly introduced and so far unknown $x$-component.
Now we know explicitly both the total cosmological dynamics, given by $w_{\mathrm{eff}}$ and the dynamics of the separately conserved matter subsystem, given by $w_{m}$. This knowledge allows us to determine the dynamics of the subsystem with the energy density $\rho_{x}$. Assuming for this component an equation of state
$p_{x} = w_{x}\rho_{x}$, it is then possible to calculate the parameter $w_{x}$.
With
$
p_{\mathrm{eff}} = p_{m} + p_{x}$ and $\rho_{\mathrm{eff}} = \rho_{m} + \rho_{x}
$
we have
\begin{equation}
w_{\mathrm{eff}} = w_{m}\frac{\rho_{m}}{\rho_{\mathrm{eff}}} + w_{x}\frac{\rho_{x}}{\rho_{\mathrm{eff}}} \quad \Rightarrow\quad w_{x} = \left(1 + r_{\mathrm{eff}}\right)w_{\mathrm{eff}} - w_{m}r_{\mathrm{eff}}
\ ,\label{w=w+w}
\end{equation}
where $r_{\mathrm{eff}} \equiv \frac{\rho_{m}}{\rho_{x}}$ is the ratio of the energy densities of both components. Using (\ref{rdiff}), we obtain for this quantity
\begin{equation}
r_{\mathrm{eff}} = \frac{\rho_{m}}{\rho_{\mathrm{eff}} - \rho_{m}} = \frac{\frac{\rho_{m}}{\rho_{\mathrm{eff}}}}{1 - \frac{\rho_{m}}{\rho_{\mathrm{eff}}}}
\ . \label{R}
\end{equation}
Different from the earlier introduced constant energy density ratio $r$ in the Einstein frame, the generally time dependent $r_{\mathrm{eff}}$ is considered to be the ratio of the energy densities of the dynamically relevant components within the effective GR description.
With the solutions (\ref{drtm2}) and (\ref{reff}) the ratio $ \frac{\rho_{m}}{\rho_{\mathrm{eff}}}$ scales as
\begin{equation}
\frac{\rho_{m}}{\rho_{\mathrm{eff}}} = \frac{\rho_{m0}}{\rho_{\mathrm{eff0}}} a^{-3\left(w_{m}-w_{\mathrm{eff}}\right)}
\ , \label{R=Ri}
\end{equation}
where $\frac{\rho_{m0}}{\rho_{\mathrm{eff0}}}$ is the present ratio of matter energy to total energy.
Consequently, for $w_{m}=0$,
\begin{equation}
w_{x} = \frac{w_{\mathrm{eff}}}{1 - \frac{\rho_{m0}}{\rho_{\mathrm{eff0}}} a^{3 w_{\mathrm{eff}}}} = \frac{1}{3}\frac{2q-1}{1 - \Omega_{m0} a^{\left(2 q -1\right)}} = \frac{1}{3}\frac{2 q -1}{1 - \Omega_{m0}\left(1+z\right)^{\left(1 - 2q\right)}}
\ , \label{wx}
\end{equation}
where $z$ is the redshift parameter and $\Omega_{m0} \equiv \frac{\rho_{m0}}{\rho_{\mathrm{eff0}}}$.

As an immediate consequence we find that in the matter era $w_{\mathrm{eff}} = 0$ one has $w_{x}=0$ as well. This behavior is reminiscent of specific interacting holographic dark energy models \cite{WDCQG}.
But $w_{x}$ is not constant for $w_{\mathrm{eff}} \neq 0$. The EoS parameter today is
\begin{equation}
w_{x0} = \frac{w_{\mathrm{eff}}}{1 - \Omega_{m0}}
\ . \label{wx+}
\end{equation}
For the present accelerated expansion, i.e., $w_{\mathrm{eff}} < 0$ one has $|w_{x0}| > |w_{\mathrm{eff}}|$, i.e., $w_{x0}$ is more negative than $w_{\mathrm{eff}}$,
since $\Omega_{m0} < 1$. This means, the $x$-component, here of geometrical origin, effectively behaves as dark energy.
Now, it is straightforward to make contact with the $\Lambda$CDM model. In the latter we have for the total EoS parameter $w_{\mathrm{LCDM}}$,
\begin{equation}
w_{\mathrm{LCDM}} = \frac{p_{\Lambda}}{\rho_{\Lambda}+\rho_{m0}}
\ . \label{wlcdm}
\end{equation}
With $p_{\Lambda} = - \rho_{\Lambda}$ and, observationally, $\rho_{m0} \approx \frac{1}{3}\rho_{\Lambda}$, the total effective EoS parameter of the $\Lambda$CDM model at the present time is $w_{\mathrm{LCDM}} \approx - \frac{3}{4}$.
Identifying the latter tentatively with $w_{\mathrm{eff}}$ in (\ref{wx}), we find $w_{x0} \approx -1$.
Consequently, the scalar-tensor-theory solution for the present epoch is compatible with the general-relativity based $\Lambda$CDM model, but the equivalent of the EoS parameter for the dark energy is time varying.
For a future attractor solution $w_{\mathrm{eff}} = -1$, one has $w_{x} < -1$. For $a\gg 1$ the parameter $w_{x}$ approaches $w_{\mathrm{eff}}$ from the phantom side.
While we have a power-law solution for the cosmic medium as a whole, corresponding to a constant EoS  parameter $w_{\mathrm{eff}}$, the EoS parameter of the geometrized dark energy is time dependent according to (\ref{wx}). The time dependence itself is again governed by $w_{\mathrm{eff}}$.

\section{Observations}
\label{observations}

As a final step we test our power-law solution against the supernovae type Ia (SNIa) data and compare the results with the corresponding analysis within the $\Lambda$CDM model.
The relevant quantity here is the luminosity distance which, in a spatially flat universe, is given by
\begin{equation}
d_L = \left(1+z\right) \int \frac{d z}{H}
\ . \label{defdl}
\end{equation}
According to (\ref{tatt}) the Hubble rate is
\begin{equation}
H = H_{0} a^{-\frac{3}{2}\left(1 + w_{\mathrm{eff}}\right)} = H_{0}\left(1+z\right)^{\frac{3}{2}\left(1 + w_{\mathrm{eff}}\right)}
\ .\label{Hobs}
\end{equation}
Then, taking into account (\ref{tilqeff}), the result of the integration is
\begin{equation}
d_L = \frac{1+z}{H_0} \frac{1}{\left(-q_{0}\right)}\left[\left(1+z\right)^{-q_{0}}-1\right]
\ , \label{dlres}
\end{equation}
i.e., the luminosity distance is exactly known. We are interested in applying this expression
around the present epoch.
Expanding in powers of $z$, we obtain for the luminosity distance up to the third order in $z$
\begin{equation}
d_L = \frac{z}{H_{0}}\left[1 + \frac{1}{2}\left(1-q_{0}\right)z + \frac{1}{6}\left(q_{0}^{2} -1\right)z^{2} + ...\right]
\ . \label{dlfin}
\end{equation}
The corresponding expression of the $\Lambda$CDM model is
(cf. \cite{visser,scaling})
\begin{equation}
d_L^{\Lambda \mathrm{CDM}} = \frac{z}{H_{0}}\left[1 +
\frac{1}{2}\left(1 -q_0 \right)z + \frac{1}{6}\left( 3\left(q_0 +
1\right)^2 - 5\left(q_0 + 1\right) \right)z^2 +
...\right]\  .
\label{dllcdm}
\end{equation}
Up to second order, the luminosity distances of both models coincide. But they differ at third order.

\begin{center}
\begin{figure}[!t]
\begin{minipage}[t]{0.25\linewidth}
\includegraphics[width=\linewidth]{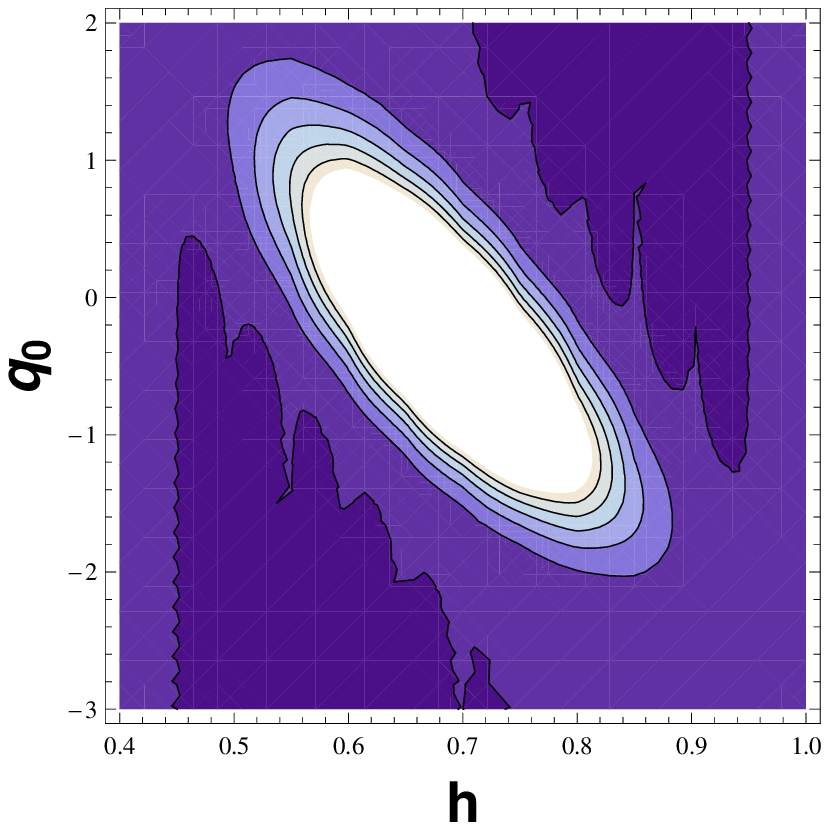}
\end{minipage} \hfill
\begin{minipage}[t]{0.25\linewidth}
\includegraphics[width=\linewidth]{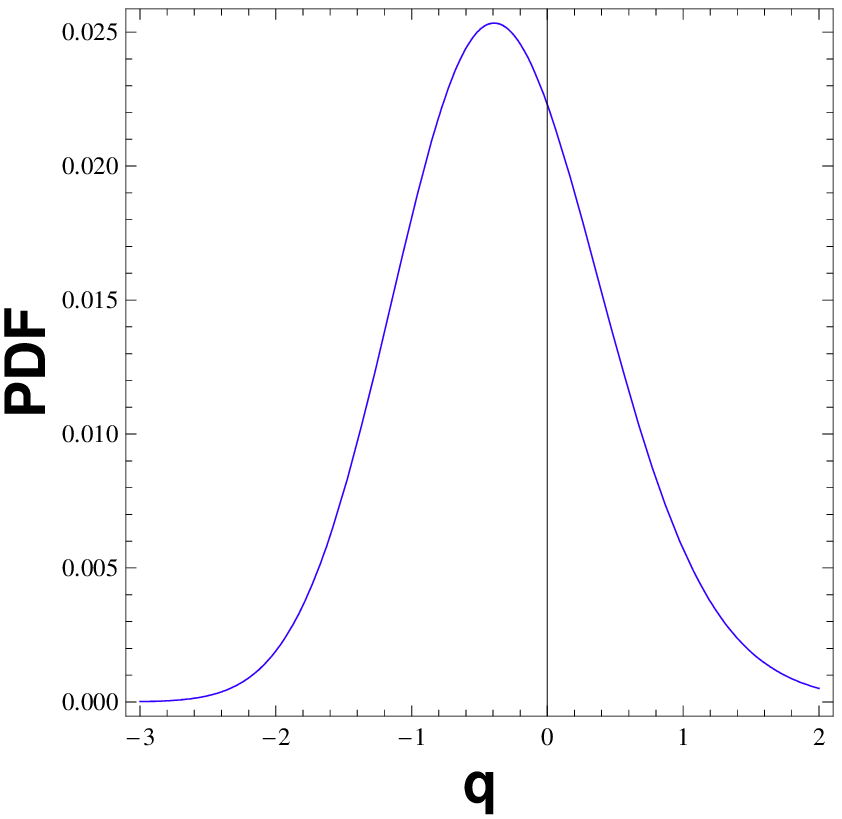}
\end{minipage} \hfill
\begin{minipage}[t]{0.25\linewidth}
\includegraphics[width=\linewidth]{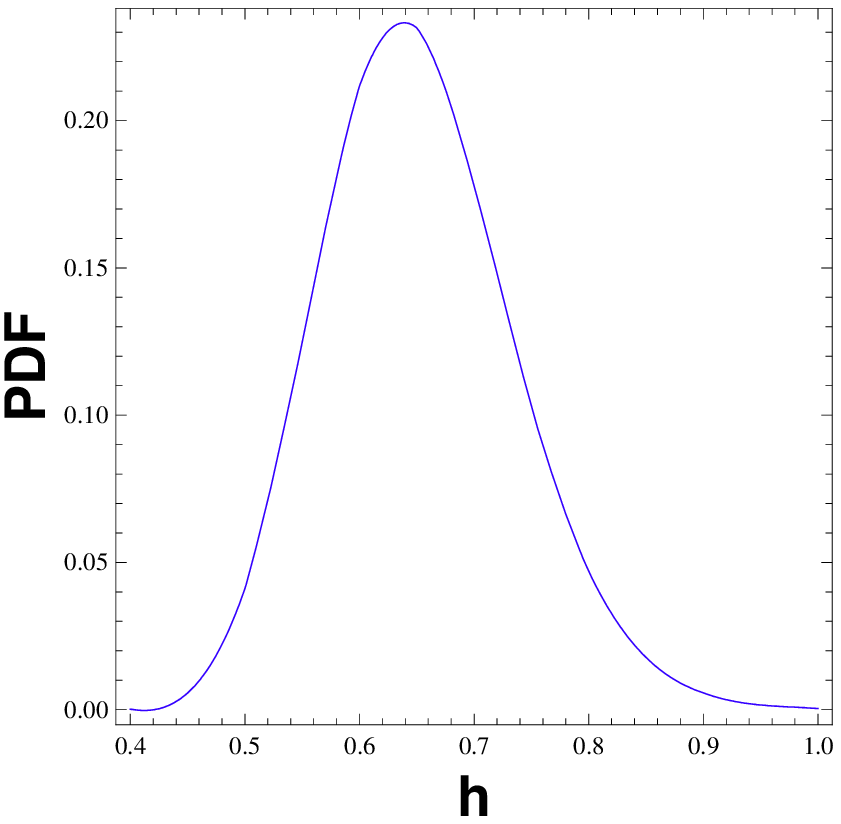}
\end{minipage} \hfill
\caption{{\label{obs}\protect\footnotesize Left panel: two-dimensional probability distribution function for $q_0$ and $h$. The brighter the color, the higher  the
probability. The best fit corresponds to $q_{0}= -0.32$ and $h = 0.64$.  Center panel: one-dimensional probability distribution function for $q_0$. The probability is maximal for $q_0 = -0.39$. Right panel: one-dimensional probability distribution function for $h$. The probability is maximal for $h = 0.64$.}}
\end{figure}
\end{center}

In the following
we will use the sample of 182 SNe Ia of
the Gold06 data set \cite{riess2006}. The
crucial quantity for our analysis is the moduli distance $\mu$,
which is obtained from the luminosity distance $d_{L}$ by
\begin{equation}
\mu =5\log\biggr(\frac{d_{L}}{Mpc}\biggl)+25\ .
\end{equation}
In order to compare the theoretical results with the observations
we perform a $\chi ^{2}$ analysis, based on the expression
\begin{equation}
\chi ^{2}=\sum_{i}\frac{\left( \mu _{0,i}^{o}-\mu _{0,i}^{t}\right) ^{2}}{%
\sigma _{\mu
_{0},i}^{2}}\quad .  \label{Chi2HSTOmegab}
\end{equation}
The quantities $\mu _{0,i}^{o}$ are the measured distance moduli
for each of the supernovae of the $182$ Gold06
SNe Ia dataset \cite {riess2006}. The $\mu _{0,i}^{t}$ are the
corresponding theoretical values and the $\sigma _{\mu _{0},i}^{2}$
represent the measurement errors (cf.
\cite{riess,riess2006}).
The probability density function (PDF) for our case with the free parameters $H_0$ and $q_{0}$  is defined by
\begin{equation}
P(H_{0}, q_{0}) = A\,e^{-\chi^2(H_{0}, q_{0})/2}\ ,
\end{equation}
where $A$ is a normalization constant.
Marginalization over one of the two free parameters will lead to
corresponding one-dimensional representations of the probability density. The details
of the analysis are given
in ref. \cite{colistete}.
The left panel of fig.~\ref{obs} shows the two-dimensional probability distribution function for $q_{0}$ and $h$. Here $h$ is defined by $H_0 = 100 h$\,km/Mpc/s. The best-fit value (minimum for $\chi^{2}$) is $\chi^{2} = 3.305$, implying $q_{0}= -0.319$ and $h = 0.640$. The corresponding values for the $\Lambda$CDM model are  $\chi^{2} = 3.297$ with $\Omega_{m0} = 0.31$, equivalent to $q_{0} = - 0.535$, and $h = 0.645$. For our power-law model, the present absolute value of the deceleration parameter is smaller than in the $\Lambda$CDM model.
The one-dimensional probability distributions for $q$ and $h$ of our model are shown in the center and right panels, respectively, of figure~\ref{obs}. The maximum values are $q_0 = -0.392$ and $h = 0.639$.

We conclude that even an approximate description of the dynamics on the basis of a power-law solution provides us with results that, at least on the background level, are consistent with observations.

\section{Discussion}
\label{discussion}

We have obtained exact power-law solutions for the dynamics of scalar-tensor theories for a constant ratio of the energy densities of the matter and the scalar-field components in the Einstein frame. Under this condition,  the parameter that describes the interaction between matter and scalar field is no longer a free parameter. The Einstein-frame solutions are then transformed into the Jordan frame. The Hubble rates in both frames do not necessarily have the same sign. E.g., an expanding solution in the Jordan frame can correspond to a contracting solution in the Einstein frame. A corresponding property holds also for the deceleration parameter which may be negative in the Jordan frame while the corresponding quantity of the Einstein frame is positive.
For pressureless matter the Jordan-frame EoS parameter  $w_{\mathrm{eff}}$ is related to its Einstein-frame counterpart $\tilde{w}_{\mathrm{eff}}$ by $w_{\mathrm{eff}} =\frac{2\tilde{w}_{\mathrm{eff}}}{1 - 3\tilde{w}_{\mathrm{eff}}}$. This solution defines the potentially possible effective EoS parameters for the cosmological dynamics as a whole.
A de Sitter-type solution $w_{\mathrm{eff}} = - 1$ is obtained for
$\tilde{w}_{\mathrm{eff}}=+1$, which corresponds to a vanishing potential term in the Einstein frame. Other relevant cases, among them a matter-dominated phase, are recovered.
The existence of exact power-law solutions with $w_{\mathrm{eff}} < - \frac{1}{3}$ in a universe filled with pressureless matter can be seen as the simplest demonstration for the possibility to describe an accelerated expansion within scalar-tensor theories without a dark-energy component.
We established an effective GR  description of the Jordan-frame dynamics and identified the geometrical  equivalent of dark energy. This component has a time dependent effective EoS, the time dependence being governed by $w_{\mathrm{eff}}$ as well. The present value of the EoS parameter for such a dark-energy simulating component
is consistent with the cosmological constant of the $\Lambda$CDM model.
But the preferred present absolute value of the deceleration parameter is smaller than its $\Lambda$CDM counterpart.
Our analysis is preliminary in the sense that it is restricted to the homogeneous and isotropic background dynamics. A perturbation analysis and a comparison with large-scale structure data will be the subject of future investigation.

\vspace{1.0cm}
\noindent
{\bf Acknowledgement:} We are indebted to J\'{u}lio Fabris for discussions and for providing us with the statistical analysis of Section \ref{observations}.  Financial support by FAPES and CNPq (Brazil) is gratefully acknowledged.

\end{document}